\documentclass[a4paper,12pt]{article}
\usepackage[utf8x]{inputenc}
\usepackage[section]{placeins} 
\usepackage{float} 
\usepackage{amsmath}
\usepackage[T1,T2A]{fontenc}
\usepackage{amsfonts}
\usepackage{graphicx}
\usepackage[colorinlistoftodos]{todonotes}
\usepackage{verbatim}
\usepackage{enumitem}
\usepackage[margin=1.0in]{geometry}
\usepackage{bbm}
\usepackage[super]{nth}
\usepackage{accents}
\usepackage{bookmark}
\usepackage{pdfsync} 

\usepackage{booktabs}

\usepackage{subfigure} 

\usepackage{threeparttable}

\usepackage{parskip}

\usepackage{indentfirst}
\newcommand{\papertitle}{Compacter networks as a defensive mechanism: How firms clustered during 2015 Financial Crisis in China}
\newcommand{\yourname}{Yujue Wang}
\newcommand{\youremail}{ywang@wisc.edu}

\usepackage{natbib}
\setcitestyle{authoryear, open={(}, close={)}}

\begin{document}

\begin{center}{\LARGE \scshape \papertitle}\end{center}
\begin{center}\vspace{0.2em} {\Large \yourname\\}{\youremail}\end{center}
\begin{center}{June, 2020}\end{center}

\vspace{50pt}

\begin{center}
\large{\textbf{Abstract}}
\end{center}

The stock market's reaction to the external risk shock is closely related to the cross–shareholding network structure. This paper takes the public information of listed companies in the A–share securities market as the primary sample to study the relationship between the stock return rate, market performance, and network topology before and after China's stock market crash in 2015. Data visualization and empirical analysis demonstrate that the return rate of stocks is related to the company's traditional business ability and the social capital brought by cross–holding. 
Several heteroscedasticity tests and endogeneity tests with IV are conducted to support the robustness. The structure of the cross–shareholding network experienced upheaval after the shock, even distorting the effects of market value, and assets holding on the return rate. The enterprises in the entire shareholding network are connected more firmly to overcome systematic external risks. The number of enterprise clusters is significantly reduced during the process. Besides, the number of newly established cross–shareholding relationships shows an outbreak, which may explain the rapid maintenance of stability in the financial system. When stable clustering is formed before and after a stock crash (rather than when it occurs), the clustering coefficient of clear clustering still has an apparent positive influence on the return rate of stocks. To sum up, the compacted network may prevent the firms from pursuing aggressive earning before the financial crisis, but would protect firms from suffering relatively high losses during and after the shock.

\thispagestyle{empty}       
\newpage                   

\setcounter{page}{1}

\section{Introduction}

Since the reform and opening–up policy has been executed, China's economy booms. With financial system reform currently deepened, the state–owned enterprises have been continuously adjusting with the long–term development project. Meantime, private enterprises have started to flourish in all directions. Nevertheless, in the context of socialism with Chinese characteristics and the status quo of prosperity, the structure of the economy is worth further exploration. As for the origin behind these phenomena, the economy will appear to a certain degree of clustering with the change of capital flow and equity structure. Moreover, the securities market's reaction to external risk impact is closely related to the cross–shareholding network structure.

This paper takes the public information of listed companies in the A–share securities market as the primary sample set to study the comprehensive relationship between the stock return rate, market performance, and network topology before and after China's stock market crash in 2015. The data includes the legal representatives of listed companies with their top ten shareholders, and the fundamental financial indicators are obtained from the quarterly financial reports of each listed company. In this paper, the shareholding relationship is applied to establish a cross–shareholding network graph, and the data is converted into a visual enterprise cluster network through the nodes of capital flow and the change in shareholding structure. The social network optimization can be implemented to divide different enterprise clusters to establish empirical analysis data sets. 

Data visualization and empirical analysis demonstrate that stocks' return rate is related to the company's traditional business ability and cross–holding social capital. Net assets, net profits, and market value will have a significant impact on stock returns. Faced with the outbreak of systemic external risks, enterprises in the whole shareholding network are more closely connected, and the number of enterprise clusters is significantly reduced. After the stock market crash outbreak in China, the network topology structure established within A–share listed companies has undergone tremendous changes, and the number of cross–shareholding relationships has shown an outbreak. The process may explain the rapid stabilization of the financial system, given the influx of external capital and the potential future dealings with these firms. 

The second part of this paper will introduce the development process of related research. The third part introduces the social network model's use to construct the scale–free network of cross–shareholding and generate the preliminary regression model. The fourth part introduces the processing and optimization of the shareholding network data, combined with acquiring other information of the stock market. The fifth part outputs the regression results and then interprets and tests the econometric model. The sixth part shows the empirical results' further analysis; the seventh part is the conclusion and related policy suggestions.

\section{Research background}
  \subsection{Research Aims}
  A variety of factors usually determines the ownership structure of listed companies in China. Regional development and investment behavior are two common ways to establish an equity relationship. Through previous studies, the spatial characteristics of economic regions can be defined: the general action path of location factors on the economy. Inspired by the social network model, this paper uses the public statistics data of listed companies in China to establish an investment relationship network. For China's A–share listed companies, the top 10 investment stakeholders' financial reports are public, and data can be extracted from static statistics. The results show that a few central state–owned enterprises (mainly banks) manipulate the whole Chinese economy to some extent, reflecting the risk transmission mechanism and the efficiency of resource allocation, which is conducive to investors and policymakers to identify the enterprises with significant influence in the network during the period of risk exposure. Besides, the construction of the relationship between listed enterprises and small–sized enterprises is helpful to the analysis of the path of social capital. The results have particular guiding significance for the primary analysis model of ownership network.

  \subsection{Relative Research}
  Regional government policies will affect professional investors' behavior and then change the ownership structure and agglomeration trend. The empirical results show that institutional investors, whether independent or domestic, will consider the enterprise level and regional industrial development policies (\cite{lin2018sustainable}). Governments also play a vital role in regional economic growth. Active investors can promote policies and create more profits and social returns by building enterprise cluster networks with higher social capital returns. The usual definition of enterprise cluster (\cite{porter1998}) is broad, which means it ignores geographical scope limitation, but does not deny the objective advantages of geographical proximity to cluster formation. The paper focuses on the network connection of the enterprises in the clusters, and its structure has a considerable impact on the generation and solidification of competitive industrial advantage (which can be measured by marginal cost).

  In the context of economic reform, the allocation of resources and the evolution of social capital will become the external manifestation of economic clustering. Misallocation of resources due to policy distortion may lead to a relatively high efficiency loss (\cite{luo2012factor}). As for measuring resource allocation efficiency, \cite{wurgler2000financial} can explain the specific relationship between financial development and efficiency improvement. The study of  \cite{pan2003spill} shows that state–owned banks' credit behavior may hurt their improvement. The dynamic panel data model can be used as a complement to illustrate the broad impact of financial institutions. The evolution of industrial cluster network has the potential to deepen gradually, and the relationship between it and technological innovation is also worth exploring (Zhou Can \& Zeng Gang, 2018).

\subsection{Innovation}
As mentioned above, with the development of primary market investment theory, enterprises' investment and financing behaviors gradually become more rational, and information asymmetry caused by regional obstacles decreases. Of course, differences in development caused by local policy support still exist. Therefore, geoeconomics cannot explain the return on social capital brought by enterprise clustering sufficiently. Therefore, this paper attempts to use the social network method to measure the influence of social capital.

Social capital is a form of capital that is different from human capital and material capital. Individuals can further obtain social capital, network external benefits such as information by establishing investment relationships (\cite{bao}). Social capital can be invested over a long period with predictable returns. For social capital rooted in the investment network, its benefits may be professional, but the exclusivity of all kinds of social capital may be gradually eliminated with the merge of the social financing network. 

This situation would be more evident in small enterprises. Specifically, large listed enterprises may vertically acquire upstream enterprises to reduce the risk of price fluctuation of raw materials or establish two–way financing relationship with complementary enterprises to cooperate and compete. Small–sized start–ups may form technical cooperation due to business dealings between common shareholders (such as individual investors or institutional investors). Therefore, social capital can partially substitute for the capital needed by other enterprises for production. For example, enterprises related firmly to state–owned enterprises may be financed at a lower capital price. This convenience can be measured by social capital.

Of course, social capital also has the nature of depreciation, so it needs to be actively maintained, which is reflected in the relative changes in investment institutions' ownership structure. The collectivity of social capital is reflected in the network externality and reflected in the dependence of social capital application on the relationship between individuals.

Concerning measuring its related network effects, economic crisis was introduced as an exogenous shock variable to measure the systemic risk caused by clustering. Systemic risk consists of the initial shock and risk contagion (\cite{Yuanjing}). When discussing the systemic risk, the price–earnings ratio, turnover rate, trades, and other aspects of the stock market can describe the change of the stock index. There are also differences in risk performance and market rescue between the Chinese stock market and the American stock market (\cite{Jianbin}). Considering the measures taken to rescue the stock market during the economic crisis in China's stock market in 2015, China Securities Finance Corporation Limited (CSF) and Central Huijin Investment Co., Ltd. (Huijin) rescued the stock market, which attracted wide attention from all walks of life. According to Wind statistics, as of September 30, 2015, by consulting the list of top 10 shareholders of listed companies, CSF and Central Huijin have appeared in the top 10 shareholders of 1,365 listed companies. The two spent a total of 1.230 billion yuan to prop up the market, buying 49\% of A–shares, covering CSI main counter, small and medium board (SME), and ChiNext stocks (GEM). As the representative of the government's rescue funds, the team symbolizes absolute authority, which undoubtedly conveys a signal of maintaining stability to the outside (\cite{xingxing}).

\section{Basic models}
\subsection{Social network model}
\subsubsection{Definition}
Firstly, the paper uses information visualization to establish shareholder investment relationships and observe the enterprise cluster structure changes before and after the stock market crash in 2015, China. With the concept of the social network, the scale–free network of A–share listed companies is also established. Scale–free networks usually conform to Pareto's Law, which indicates that most nodes have few connections, while a few nodes have connections with many other nodes. Due to the critical hub nodes' connection characteristics, the scale–free network has overall stability and can withstand partial failures or shocks. However, due to the extensive and close connections among nodes, the network is more sensitive to coordinated attacks, and the impact range is more extensive. The investment network in the financial market is a typical example of a scale–free network.

\subsubsection{Network construction}
First, A–share listed companies as network nodes are defined, so three–node relationships are further generated. The first relationship is the cross–level inter–enterprise connection, which indicates that there is listed company B among the ten shareholders of a listed company A. Such a relationship exists in a small number, but it has a significant influence because the companies are A–share listed companies and greatly influence other listed companies' decision–making. The second kind of relationship is the enterprise relationship connected by individuals. That is to say, X, the ten shareholders of a listed company A, is the legal representative of another listed company. The third relationship is the relationship between listed companies of the same level, which means listed company A and listed company B are controlled by the same ten shareholders: listed companies, other enterprises, institutional investors, or even individual investors. This type of relationship is the most crucial business connection, but it is also the weakest. In fact, in general, in the network communities, the degree of a few nodes (the number of node connections) and the strength of nodes (the degree of nodes in the weighted network) are large, while the degree and the strength of nodes of most nodes are small. This feature is also reflected in the model statistics.

The cross–shareholding among listed enterprises is undoubtedly the most intuitive equity network connection. The legal representative of a listed company is relatively stable for a long time. As the only natural person who can represent all kinds of external behaviors of the listed company (that is, the personified legal person in the narrow sense), the legal persons of listed firms are the main person in charge of exercising official powers of the legal representatives. Therefore, to some extent, the legal person can replace the listed company's node position in the equity network. However, the directors and supervisors in listed companies do not have a standardized template. Furthermore, the directors and supervisors need to maintain absolute independence, but the legal representative can only be created between the chairman of the board, the executive director, or the manager. This kind of uncertainty and complicated relationship is not conducive to extracting data to construct the shareholding network with apparent influence. Therefore, the second relationship's definition is rooted in replacing the corresponding listed company nodes in the equity network with the legal representative of the listed company, and subsequent data also show that these two relationships are similar in magnitude.

Although the first and second kinds of relationships are relatively direct, the overall network construction mainly depends on the third kind of relationship. The third type of relationship is complicated, large in number, and weak in connection. However, the emergence of common shareholders is likely to come from institutional investors or individuals who focus on a specific field, vertical mergers and acquisitions in the industry chain to reduce costs, and horizontal mergers and acquisitions to obtain excess profits. Thus, the third type of connection helps the division and construction of network clusters on the whole. Unlike the former two types of relationships, the third type of relationship needs a lot of data support to optimize the network cluster.

The above three connections belong to edge attributes, which are only used for relevant network data statistics and basic network configuration construction, as the underlying logic of the shareholding model increases the richness of connections. The cross–sectional data used in the regression were listed by individual listed companies, while the edge attributes do not appear in the regression.

\subsection{The regression model}
In addition to the cross–section data indicators obtained from the social network model, the paper also uses regression test to confirm that network characteristics such as social capital, the number of investment and financing relationships, the size and stickiness of the cluster brought by network clustering can significantly affect the return rate of stocks.

Considering that systemic risk will increase exponentially with the cluster's closeness, in such an interconnected financial network, if one enterprise is represented by a node fault or goes bankrupt, all related enterprises would be affected. Financial risks are transmitted to other institutions in the system through cross–shareholding and interbank business. When the losses of institutions and systems infected and bankrupt reach a certain level, systemic risk events will break out, causing a massive impact on national finance and making it difficult to maintain the real economy's stability. In the process of systemic risk contagion, the loss of the bankruptcy of an institution to other institutions in the system depends on its owner's equity and unliquidated interbank liabilities. The loss of owner's equity is transmitted through cross–shareholding channels, and the unliquidated interbank liabilities are transmitted through interbank business channels (Yang Yuanjing, 2017). In the listed companies' financial report, the owner's equity is net assets, while interbank business transactions are related to the node degree. In the face of the sudden financial crisis and the deterioration of the overall market, large–scale institutional bankruptcy will lead to the bankruptcy of an excessive number of other financial institutions, so the risk contagion hazard presents an exponential growth trend. Therefore, a simulation model of A–share listed companies can be established, and the structural changes of economic clusters can be compared before and after China's stock market crash in 2015.

The cohesiveness of the financing network is reflected in the return rate on social capital, but its vulnerability to systemic risk also exists. The relationship between the stickiness and intensity of the financing relationship and other financial indicators is used to measure stocks' performance in the financial market. The comparison before and after the stock market crash proves that the social capital generated by enterprise clustering will aggravate the risk contagion effect, which provides empirical evidence for the formation of adverse effects of social capital from the perspective of risk, and deepens the theoretical research on the harmful effects of social capital.

\section{Data processing}
\subsection{The shareholding networks}

The data used in the shareholding network is the shareholder information of listed companies in the company research series in the CSMAR database. The top ten shareholders of non–tradable shares of each listed company are derived. The data is updated with the quarterly reports. According to the three relationships described above, the data processing is carried out as follows:

Type 1: If B, one of the top ten shareholders of company A, also belongs to the data set of the listed company, establish A$ \rightarrow $B vector.

Type 2: If I, the top ten shareholders of company A, is the legal representative of a listed company B, the vector from A to B is established.

Type 3: If the top ten shareholders of company A are also the top ten shareholders of company B, establish A $ \rightarrow $ B vector. This weak relationship is double–counted here.

After the whole data set of listed companies is traversed, node vector documents of corresponding types can be obtained in different periods. Three types of vectors are integrated as the original data sets for constructing the visualized network for each period. The dataset consists of three columns, including a starting point column, an ending point column, and a vector type annotation.

Considering that China's stock market crash in 2015 occurred around May, the selected time window is 03.01.2015–05.31.2015 (Group1), 06.01.2015–08.31.2015 (Group2), and 09.01.2015–11.30.2015 (Group3). The statistical data are shown in Table \ref{sumnw}.

\begin{table}[H]
  \centering
  \scalebox{1}{
  \begin{threeparttable}
  \caption{Cross–shareholding Connections Summary}
\label{sumnw}
\begin{tabular}{lllll}
\toprule
Group & Total connections & Type 1 & Type 2 & Type 3  \\ \hline
1     & 151386            & 359    & 275    & 150752  \\
2     & 165897            & 379    & 330    & 165188  \\
3     & 1048575           & 389    & 303    & 1047883\\
 \bottomrule
\end{tabular}
\end{threeparttable}  }     
\end{table}

Obviously, after the stock market crash, the third type of connections increased rapidly, and the growth rate was equivalent to the magnitude of 13–18 years, a sevenfold increase. The common ten shareholders link many listed companies. Due to the frequent occurrence of market rescue policies and equity transactions in the depressed stock market, the market was in turmoil. Meanwhile, the structure of the cross–shareholding network changed on a large scale. Besides, the number of groups dropped sharply after June, reflecting more diversified but more quantitatively linked investment relationships. The phenomenon could result from diversified investment, such as a surge in funds investing in the same listed company.

\subsection{Network optimization}
\subsubsection{Network topology statistics}
According to the vector file developed in 4.1, Gephi software is used to realize the network's establishment and optimization.

First, calculate the model's degree (including in–degree and out–degree). The degree means the number of connections in and out of each node within the directed network, which are simplified as edges. Then, the Force Atlas2 optimization method, a force–oriented mass layout, was used for all three groups. The method can be thought of as a repulsive linear model combined with simulation algorithms like Barnes Hut. The next step is to find and categorize the communities by classifying nodes according to the graph's connections, which are called modularization. By optimizing the modularized distribution, the value was minimized. Under the probability of 95\%, the three groups were divided into 9, 8, and 3. However, in the actual regression, all the modular groups are still adopted as dummy variables. As a general rule of thumb, the modularization process would be significant if Modularity Coefficient is greater than 0.4. Also, although there is almost no node change, which means the number of delisted enterprises is 1, the situation of delayed listing due to the stock market crash exists compared with the increase in the last quarter.

In addition to the significant increase in the number of edges, the network diameter also becomes larger after the stock market crash. The increase of the average clustering coefficient indicates that the whole network is more closely linked. The number of nodes near each node increases compared with that before. Correlative to the index of clustering compactness, the average path length is also shortened. The fact shows that in the stock market crash process, the whole network has contracted. However, due to overcrowding, the significant modularization category is reduced to 3, and the modularization index is reduced.

\begin{table}[H]
  \centering
  \scalebox{1}{
  \begin{threeparttable}
  \caption{Statistical characteristics of cross–shareholding networks}
\label{stat}
\begin{tabular}{lllll}
\toprule
                                     & 1               & 2               & 3               \\\hline
Node                                 & 2609            & 2720            & 2719            \\ 
Edge                                 & 141644          & 152428          & 1207363         \\
Average degree                       & 54.291          & 56.04           & 444.047         \\
Network diameter                     & 7               & 7               & 8               \\
Average clustering coefficient       & 0.456           & 0.454           & 0.682           \\
Average path length                  & 2.686           & 2.698           & 2.165           \\
Modularity Coefficient               & 0.422           & 0.451           & 0.217           \\
Modularity class (\textgreater{}5\%) & 9               & 8               & 3               \\
Algorithm                            & Force   Atlas 2 & Force   Atlas 2 & Force   Atlas 2\\
 \bottomrule
\end{tabular}
\end{threeparttable}  }     
\end{table}

The module size of each group is shown in Figure \ref{sizedis}, indicating the number of nodes in each modular category:

\begin{figure}[H]
  \centering
\caption[]{Cluster Size Distribution}
  \subfigure[Group 1]{
  \begin{minipage}[t]{0.33\linewidth}
  \centering
  \includegraphics[width=2in]{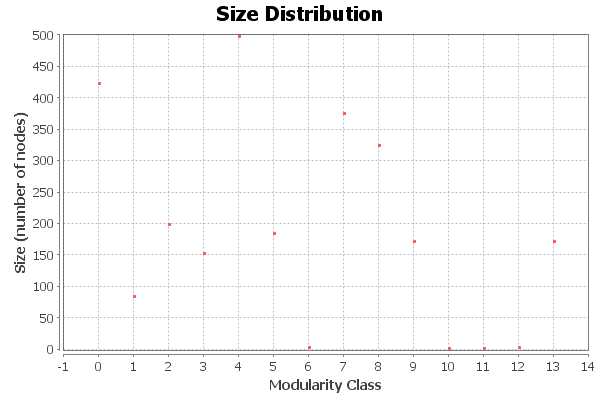}
  \end{minipage}%
  }%
  \subfigure[Group 2]{
  \begin{minipage}[t]{0.33\linewidth}
  \centering
  \includegraphics[width=2in]{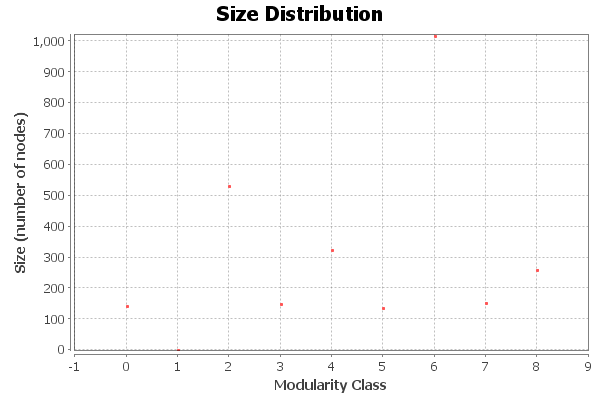}
  \end{minipage}%
  }%
  \subfigure[Group 3]{
  \begin{minipage}[t]{0.33\linewidth}
  \centering
  \includegraphics[width=2in]{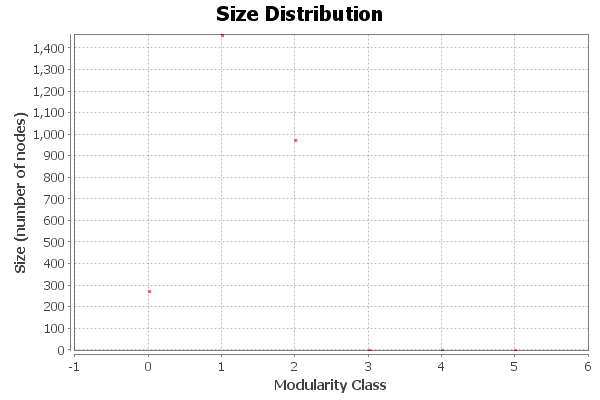}
  \end{minipage}%
  }%

  \label{sizedis}
\end{figure}

The degree distribution of each module in each data set is shown in Figure \ref{id}, including the number of nodes with the same in–degree, out–degree, and degree:

\begin{figure}[H]
  \centering
\caption[]{Degree Distribution}
  \subfigure[Group 1]{
  \begin{minipage}[t]{0.33\linewidth}
  \centering
  \includegraphics[width=2in]{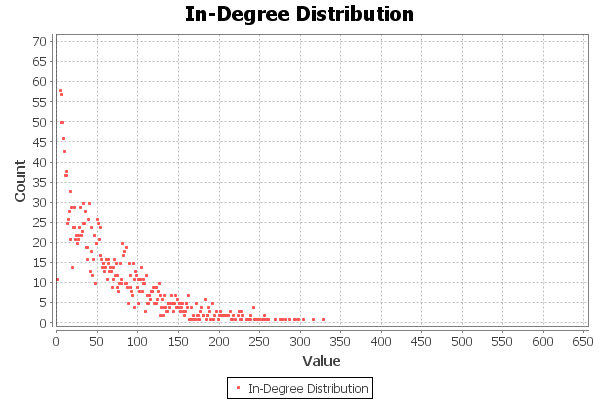}
  \end{minipage}%
  }%
  \subfigure[Group 1]{
  \begin{minipage}[t]{0.33\linewidth}
  \centering
  \includegraphics[width=2in]{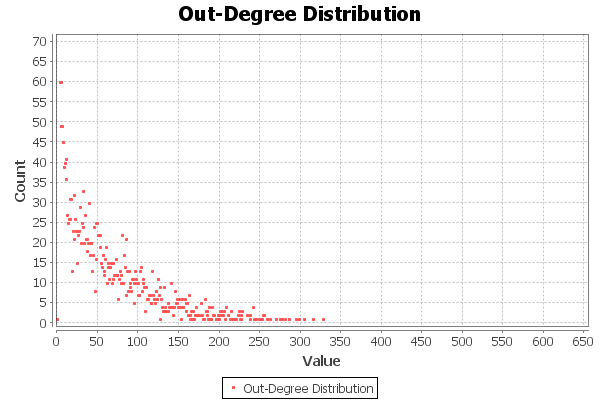}
  \end{minipage}%
  }%
  \subfigure[Group 1]{
  \begin{minipage}[t]{0.33\linewidth}
  \centering
  \includegraphics[width=2in]{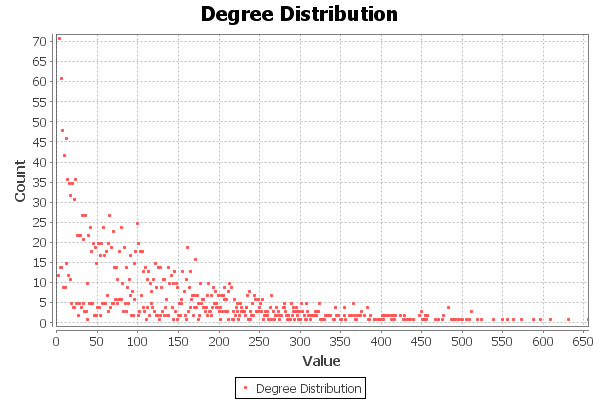}
  \end{minipage}%
  }%

  \subfigure[Group 2]{
    \begin{minipage}[t]{0.33\linewidth}
    \centering
    \includegraphics[width=2in]{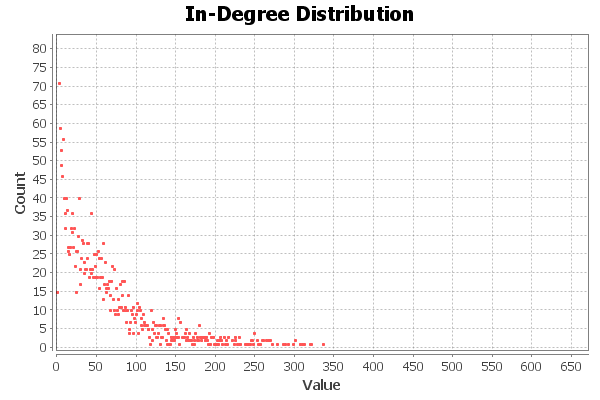}
    \end{minipage}%
    }%
    \subfigure[Group 2]{
    \begin{minipage}[t]{0.33\linewidth}
    \centering
    \includegraphics[width=2in]{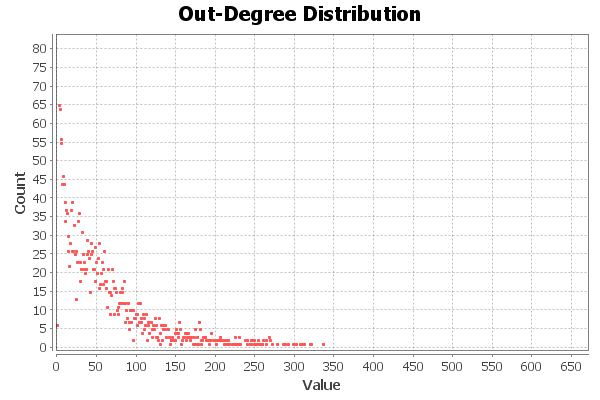}
    \end{minipage}%
    }%
    \subfigure[Group 2]{
    \begin{minipage}[t]{0.33\linewidth}
    \centering
    \includegraphics[width=2in]{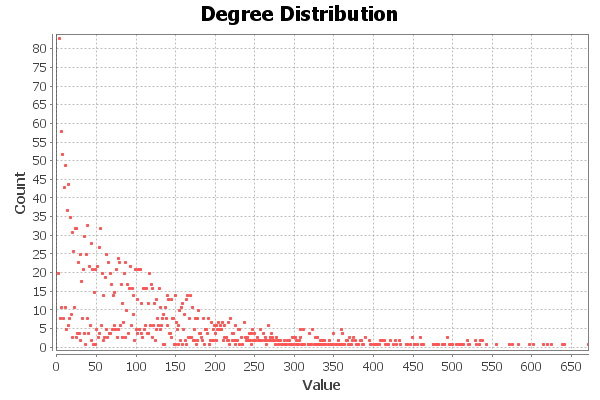}
    \end{minipage}%
  }%

    \subfigure[Group 3]{
      \begin{minipage}[t]{0.33\linewidth}
      \centering
      \includegraphics[width=2in]{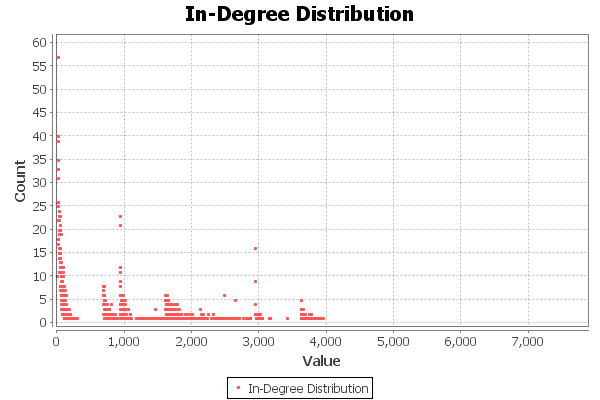}
      \end{minipage}%
      }%
      \subfigure[Group 3]{
      \begin{minipage}[t]{0.33\linewidth}
      \centering
      \includegraphics[width=2in]{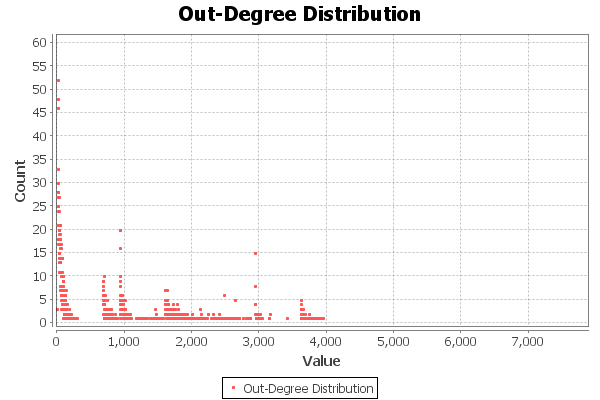}
      \end{minipage}%
      }%
      \subfigure[Group 3]{
      \begin{minipage}[t]{0.33\linewidth}
      \centering
      \includegraphics[width=2in]{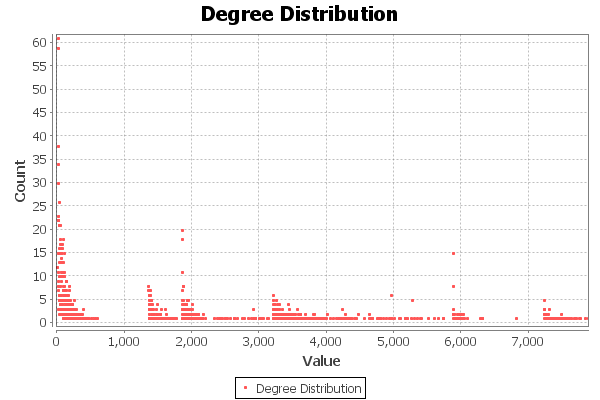}
      \end{minipage}%
  }%

  \label{id}
\end{figure}  

\begin{figure}[H]
  \centering
\caption[]{Equity Network Visualization }
  \subfigure[Group 1]{
  \begin{minipage}[t]{0.33\linewidth}
  \centering
  \includegraphics[width=2in]{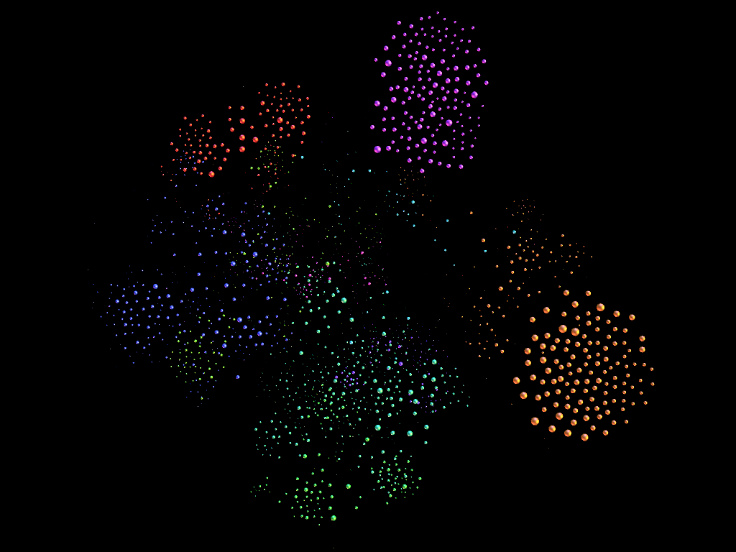}
  \end{minipage}%
  }%
  \subfigure[Group 2]{
  \begin{minipage}[t]{0.33\linewidth}
  \centering
  \includegraphics[width=2in]{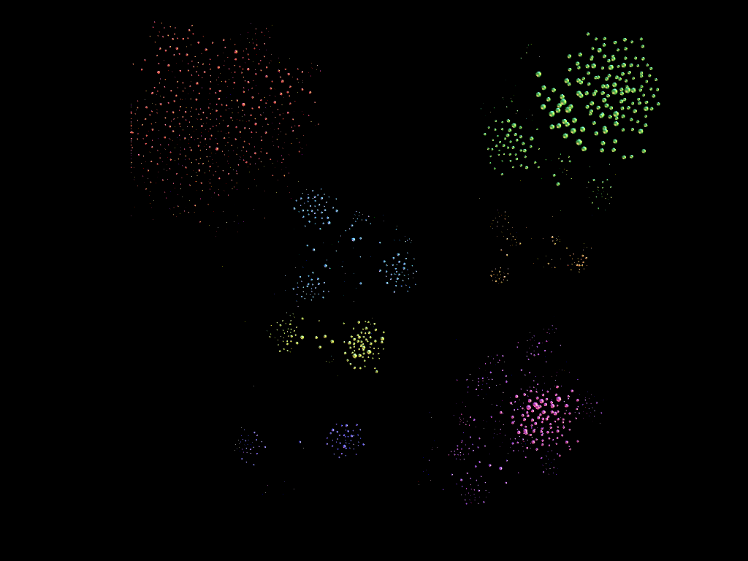}
  \end{minipage}%
  }%
  \subfigure[Group 3]{
  \begin{minipage}[t]{0.33\linewidth}
  \centering
  \includegraphics[width=2in]{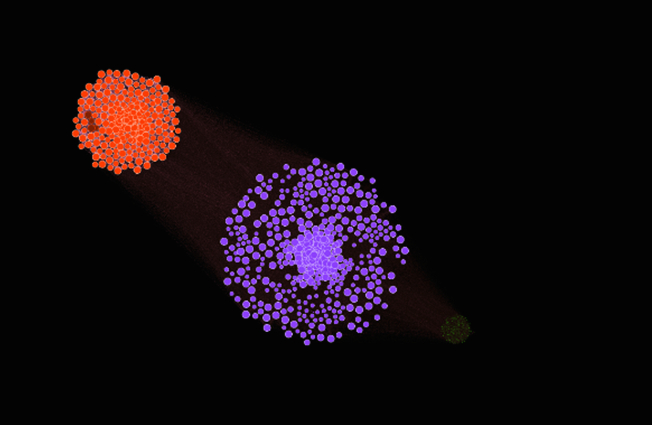}
  \end{minipage}%
  }%

  \label{vis}
\end{figure}

The frequency distribution of observed degrees is shown in the figure above. Obviously, before the stock market crash, the distribution is similar to the exponential distribution. After the stock market crash, faults appear in its image, as shown in Fig \ref{vis}.

Through model optimization, the enterprise network topology diagrams of the three data sets are optimized respectively. The above schematic diagram is obtained, in which different colors represent different module sets, and the size of nodes is simulated according to the degree of nodes to show the influence.

\subsection{Parameter output}
After the optimization, the Centrality indicator calculates the average path length and network diameter, including Betweenness Centrality, Closeness Centrality, Eigenvector Centrality, and Eccentricity. Betweenness centrality refers to the number of the shortest paths between all node pairs that pass through this node. The greater betweenness centrality means that it is a hot hub node capable of connecting more data nodes. Closeness centrality is the sum of the number of reachable nodes divided by the shortest path of reachable nodes. The location advantage of this type of node is that it has the highest accessibility. The eigenvector centrality means that this class of nodes' centrality depends on the centrality of neighboring nodes. The eccentricity is the shortest path to the farthest point a node can go.

The Clustering Coefficient  is also calculated, which can represent the overall indication of some nodes' clustering and classification, and measure a single node's neighborhood integrity. The neighborhood of a node is the set of nodes connected to the node. If each node in this central node's neighborhood is connected to any node in the neighborhood, then the node's neighborhood is complete, and the clustering coefficient is 1. If there is no node connection in the neighborhood of the node, the clustering coefficient is 0.

\subsection{Stock market performance}
Given the performance of stocks in the financial market, this paper selects the monthly return rate of stocks of listed companies (the monthly return rate of stocks with the reinvestment of cash dividends is adopted in this paper), the total market value of stocks, and the monthly circulation of stocks from the CSMAR database. These indicators are averaged according to the previously selected groups to obtain quarterly indicators.

Then, the owner's equity (net assets) and net profit of all companies in the current quarter are extracted from listed companies' financial statement database. According to the above indicators, PE, PB, ROE, and other financial indicators can be further obtained.

A stock symbol connects the above data, and null values are removed during filtering. The data can be used for empirical analysis based on stocks' performance in China's A–share market. By running and optimizing the network topology diagram and obtaining the market performance of stocks, primary data can be obtained to support the empirical analysis. In addition to the traditional variables such as net assets and the market value of listed companies, the paper attempts to add variables such as node degree and network topology centrality. It verifies the effectiveness of these new variables for estimating the return rate of stocks.

\section{Regression model}
\subsection{Econometric model}
According to the previous data introduction, the econometric model in this paper is as follows:
\begin{equation}
  Y_i=log(V_i)+log(NPP_i)+NPF_i+TD_i+Degree_i+Centrality_i+\Sigma_mDm*MC_{mt} \label{reg1}
  \end{equation}

  Here $Y_i$ stands for the quarterly average monthly return on stocks taking into account the reinvestment of cash dividends. Vi represents the market value of the listed company in the current period. $NPP_i$ (NET\_PROPERTY) is the net assets of the current period. $NPF_i$ (NET\_PROFIT) is the net profit of the current period. TD\_i is the quarterly average monthly value of stock trades. Degree\_i is the node degree of the listed company. Centrality\_i is a set of variables about node centrality. DM is m dummy variables used to group listed companies according to modularization, and M is the number of modularization groups.

This model will be applied to three periods before and after the stock market crash in 2015, and the impact of network topology changes on the explanatory ability of clustering on stock returns will be compared.

\subsection{Regression results}

Due to the extensive sample cross–section data, OLS robust standard error method was adopted in the regression.

The following equations are used in the actual regression:
\begin{equation}
  Y_i=log(V_i)+log(NPP_i)+NPF\_d_i+NPF\_d_i^2+Degree_i+Centrality_i+\Sigma_{m–1}Dm*MC_{mt}\label{reg2}
  \end{equation}

Where $NPF\_d_i$ is the standardized net profit of the current period, and $NPF\_d_i^2$ is the quadratic term of the current net profit. In this regression, the quarterly average monthly trading amount of stocks was excluded because its regression coefficient's t–value was so large to affect the Equation's regression validity. Meanwhile, $Centrality_i$ variable used Betweenness centrality and Clustering coefficient. To avoid the trap of dummy variables, only m–1 dummy variables are selected, where m is the number of the modular groups.

The results of robust standard error regression are shown in Table \ref{reg}:

\begin{table}[H]
  \centering
  \begin{threeparttable}
  \caption{Regression, robust in standard error}
\label{reg}
\begin{tabular}{llll}
\toprule
     & 2015.03.01–2015.05.31 & 2015.06.01–2015.08.31 & 2015.09.01–201511.30 \\ \hline
log(V\_i)                 & 0.0866***             & 0.03461***            & 0.03137***           \\
                          & –0.01378              & –0.006696             & –0.0066              \\
log(NPP)                  & –0.07861***           & –0.0150***            & –0.0515***           \\
                          & –0.008968             & –0.005028             & –0.00512             \\
NPF\_d                    & –0.0236***            & 0.00017               & 0.01851***           \\
                          & –0.00903              & –0.0053045            & –0.004094            \\
NPF\_d\textasciicircum{}2 & 0.001037***           & 0.00011               & –0.00049***          \\
                          & –0.0004               & –0.00021              & –0.000143            \\
Degree                    & –0.00037***           & –0.00019***           & 0.000015**           \\
                          & –0.0000961            & –0.0000492            & –0.00000711          \\
Betweenness Centrality    & 0.00000293**          & 0.00000132*           & –0.000000321         \\
                          & –0.00000159           & –0.000000834          & –0.000000591         \\
Coefficient               & 0.04662*              & 0.0095                & 0.03245*             \\
                          & –0.02895              & –0.01859              & –0.0173              \\
Const                     & 0.5802***             & –0.3223***            & 0.7021***            \\
                          & –0.02598              & –0.075675             & –0.060469            \\
Modularity\_Class         &                       &                       &                      \\
1                         & –0.0581296***         & –                     & –0.00022             \\
2                         & –0.0368772            & –0.0203               & 0.002438             \\
3                         & –0.0409863**          & –0.00943              & –0.1094              \\
4                         & –0.0630825***         & –8.40E–05             & –0.07124***          \\
5                         & –0.0292891            & –0.00428              & 0.004463             \\
6                         & 0.7951872             & –0.04168***           &                      \\
7                         & –0.0109385            & –0.05003***           &                      \\
8                         & –0.0606918***         & –0.03208**            &                      \\
9                         & –0.0691154***         &                       &                      \\
10                        & –0.2209544***         &                       &                      \\
11                        & –0.0930061            &                       &                      \\
12                        & 0.0004591             &                       &                      \\
13                        & –0.0456537*           &                       &                      \\
R–Square                  & 0.1139                & 0.0235                & 0.1199               \\
F–value                   & 14.37***              & 6.56***               & 50.3***              \\
Sample size               & 2,540                 & 2568                  & 2547           \\ 

\bottomrule
\end{tabular}
\begin{tablenotes}
\footnotesize
\item[]
Standard errors are in parentheses.\\
$ ^{***} p<.01, ^{**} p<.05, ^* p <.1$\\
This regression is an estimate of Equation \eqref{reg2}. In the second group, the group with Modularity class of 1 was deleted because only one variable could not be estimated.
\end{tablenotes}        
\end{threeparttable}       
\end{table}

The three regression groups' R–squares are generally small because there are many independent variables, and their distribution is relatively dense, which indicates that the change of independent variables is more evident than the change of fitted values. Considering that the t–tests of the independent variables pass, multicollinearity is excluded.

After the BP test, there is a heteroscedasticity problem in the sample. Given this problem, the robust standard error method, WLS method, and WLS method with robust standard error are applied to re–estimate the regression. Finally, the robust standard error method is used to estimate the non–zero significance of the t–test of each explanatory variable and the F–test of the Equation.

The first group has significant coefficients in terms of market performance, network clustering, and modular grouping. The fitting of variables related to the net income in the second group's market performance is not good enough. According to the estimated modularization coefficient, the modularized groups account for 53.17\% of the total number of nodes, with a significant probability of more than 99\%. The third group has a good fit in terms of network cluster and market performance, while the number of modularized groups is small, indicating that the grouping effect is poor.

\subsection{Heteroscedasticity and endogeneity tests}
\subsubsection{Heteroscedasticity tests}

The first group is verified that the residuals are highly correlated with the dependent variable (correlation coefficient is 0.9413), but not with other independent variables (correlation coefficients are all 0). Thus, the estimation is unbiased. Therefore, the RESET test is used to determine whether there are missing variables, which shows F(20,2500)=4.48. Therefore, the null hypothesis of no missing variables is rejected, and the heteroscedasticity is admitted, which is consistent with the conclusion obtained by the BP test of fitted values and explanatory variables, respectively. BP test on the fitted value shows that chi2 (1)=70.67, and BP test on the independent variable shows that chi2 (20)=98.94. Both tests reject the null homoscedasticity hypothesis, so the heteroscedasticity problem is admitted.

In the second group, the residual is highly correlated with the dependent variable (the correlation coefficient is 0.9882) after estimation, but not with other independent variables (the correlation coefficients are all 0). So the estimation is unbiased. RESET test is also used for missing variables, which showed F(20, 2534)=1.71. There is more than a 95\% probability to reject the null hypothesis of no missing variables and admit the existence of heteroscedasticity, which is consistent with the conclusion obtained by the BP test of fitted values and explanatory variables, respectively. BP test on the fitted value shows that chi2 (1)=12.87, and BP test on the independent variable shows that chi2 (14)= 48.72. Both tests reject the homoscedasticity null hypothesis and remind the heteroscedasticity problem.

In the third group, the residual is highly correlated with the dependent variable (the correlation coefficient is 0.9381) after estimation, but not with other independent variables (the correlation coefficients are all 0). So the estimation is unbiased. The RESET test is also used for missing variables, which shows F(20, 2515)=6.12. Therefore, the null hypothesis of no missing variables is rejected, and heteroscedasticity is admitted, consistent with the conclusion obtained by the BP test of fitted values and explanatory variables, respectively. BP test on the fitted value shows that chi2 (1)=22.98, and BP test on the independent variable shows that chi2 (14)= 34.06, both of which reject the null homoscedasticity hypothesis and admit the heteroscedasticity problem.

Therefore, in estimating the last part, the robust standard error method is adopted to estimate the three groups because the sample size collected in this paper is enormous. Even in the case of heteroscedasticity, the robust standard error method could help complete all parameter estimation and hypothesis testing without bias.

\subsubsection{Endogeneity tests}
The first and third groups perform well in the fitting, but the second group does not fit well on the market performance. Markets are functioning smoothly before the crash. After the rescue is complete, the close links between the listed companies will also suddenly form. However, when the stock market crash occurs and the system fluctuates wildly, the model's prediction ability will be significantly reduced. If the model is exposed to drastic fluctuations due to unknown external financial risks, there may be severe endogeneity problems.

Due to the drastic changes in the cross–shareholding network, the financial market fluctuations, such as market confidence and capital circulation, may be included in the residual changes. Therefore, the estimated value of net profit may be ineffective because stocks' net profit is related to the fluctuations in the financial market. Therefore, in the endogeneity test, the quarterly average monthly trading amount of stocks is introduced as the instrumental variable. The trading amount of tocks can be used to calculate the turnover rate and other stock market indicators. It is an essential measure of stock liquidity, so it is also related to the net profit. However, since this variable comes from the summary information of stock market trading, it is not directly related to network topology characteristics. The correlation coefficient matrix with other variables also supports this view (see Table \ref{corrmat}). The correlation coefficient between the trading amount of stocks and the logarithmic variable of net profit is large and equal to 0.4154. However, the correlation coefficient between stock trading amount and the residual of robust standard error regression is –0.0337, with a small absolute value. Besides, although the trading amount of stocks is related to substantial changes in equity, it is also related to individual investors' investment behavior. So this variable can be used as an appropriate instrumental variable.

\begin{table}[H]
  \centering
  \scalebox{0.5}{
  \begin{threeparttable}
  \caption{Pairwise correlations matrix of variables}
\label{corrmat}
\begin{tabular}{lllllllllllll}
\toprule
& Mretwd  & log\_ppt & log\_v  & NPT\_d  & mnvaltrd & degree  & Eigenvector  & Closeness  & Betweenness  & Eccentricity & Clustering Cofficient & Residual \\ \hline
Mretwd                  & 1       &          &         &         &          &         &                        &                      &                        &              &                      &          \\
log\_ppt                & 0.0334  & 1        &         &         &          &         &                        &                      &                        &              &                      &          \\
log\_v                  & 0.0827  & 0.8032   & 1       &         &          &         &                        &                      &                        &              &                      &          \\
NPT\_d                  & 0.0335  & 0.3786   & 0.347   & 1       &          &         &                        &                      &                        &              &                      &          \\
mnvaltrd                & 0.0212  & 0.6121   & 0.6851  & 0.4154  & 1        &         &                        &                      &                        &              &                      &          \\
degree                  & –0.0252 & 0.3835   & 0.3915  & 0.1613  & 0.3334   & 1       &                        &                      &                        &              &                      &          \\
Eigenvector\_Centrality & 0.0053  & 0.4301   & 0.4163  & 0.2107  & 0.3827   & 0.8809  & 1                      &                      &                        &              &                      &          \\
Closenness\_Centrality  & 0.0926  & –0.2704  & –0.2628 & –0.0821 & –0.1997  & –0.7318 & –0.5065                & 1                    &                        &              &                      &          \\
Betweenness\_Centrality & –0.0169 & 0.2181   & 0.2252  & 0.0653  & 0.1935   & 0.794   & 0.574                  & –0.5761              & 1                      &              &                      &          \\
Eccentricity            & 0.0606  & –0.293   & –0.2737 & –0.105  & –0.2252  & –0.6683 & –0.6098                & 0.7694               & –0.4941                & 1            &                      &          \\
Clustering Coefficient    & 0.0117  & 0.0424   & 0.0134  & 0.0298  & 0.0288   & 0.0773  & 0.1715                 & –0.1644              & –0.1447                & –0.1523      & 1                    &          \\
Residual                & 0.9882  & 0        & 0       & 0       & –0.0337  & 0       & 0.0116                 & 0.0585               & 0                      & 0.0432       & 0                    & 1    \\
\bottomrule
\end{tabular}        
\end{threeparttable}      }
\end{table}

Thus, the following model is established:

\begin{align}
  NPF\_d_i=&\alpha_0 +\alpha_1 *Mnvaltrd_i  \tag{3a}\\
  Y_i=&log(V_i)+log(NPP_i )+NPF\_d_i  \notag\\
  &+Degree_i+Centrality_i+∑_(m–1)D_m*MC_mt +const    \tag{3b}\\
\end{align}

OLS regression and 2SLS instrumental variable methods are implemented respectively to develop the results below:

\begin{table}[H]
  \centering
  \begin{threeparttable}
  \caption{2SLS Regression}
\label{2sls}
\begin{tabular}{lll}
\toprule
& OLS                  & 2SLS                     \\ \hline
log(NPP)                  & –0.01502832***       & –0.007792                \\
                          & –0.00502836          & –0.005987                \\
log(V)                    & 0.03460819***        & 0.03920***               \\
                          & –0.00669584          & –0.007744                \\
NPF\_d                    & 0.00016908           & –0.03347**               \\
                          & –0.00530446          & –0.0153                  \\
NPF\_d\textasciicircum{}2 & 0.00010655           &                          \\
                          & –0.00020715          &                          \\
Degree                    & –0.00019117***       & –0.0001594***            \\
                          & –0.0000492           & –0.0000526               \\
Betweenness Centrality    & 0.00000132*          & 0.000000786              \\
                          & –0.000000834         & –0.000000893             \\
Clustering Coefficient    & 0.0095               & 0.00913                  \\
                          & –0.01859             & –0.0187                  \\
Const                     & –0.3223***           & –0.5537***               \\
                          & –0.075675            & (–0.1243)                \\
Modularity Class          &                      &                          \\
1                         & –                    & –                        \\
2                         & –0.0203              & –0.01757                 \\
3                         & –0.00943             & –0.0003777               \\
4                         & –8.40E–05            & –0.000176                \\
5                         & –0.00428             & –0.00172                 \\
6                         & –.04168168***        & –.04071***               \\
7                         & –.05003159***        & –.050621***              \\
8                         & –.03207691**         & –.03323**                \\
R–Square                  & 0.0235               & 0.2028(1st stage)        \\
Significance              & F(14, 2553) =6.56*** & Wald chi2(13) = 61.93*** \\
Sample Size               & 2,568                & 2568                       \\  

\bottomrule
\end{tabular}
\begin{tablenotes}
\footnotesize
\item[]
Standard errors are in parentheses.\\
$ ^{***} p<.01, ^{**} p<.05, ^* p <.1$\\
The OLS regression is an estimate of Equation \eqref{reg1}. The 2SLS regression estimates Equation (3a) and Equation (3b). 
\end{tablenotes}        
\end{threeparttable}       
\end{table}

Hausman test for heteroscedasticity, i.e. robust heteroscedasticity DWH test shows that, robust score chi2(1)=10.3823(p = 0.0013), and robust regression F(1,2553)=11.8211(p = 0.0006). Thus, there are endogenous variables in this regression.

\subsection{Analysis}

Similar to the prediction, the nature of network topology has a significant impact on the return rate of stocks, and its influence coefficient is still significant even in the particular period in the face of the stock crisis.

The model takes a logarithmic approach to net assets and total market values, consistent with the marginal increase in capital return. The results show that the market value has a significant positive effect on the return rate of stocks. Each unit increase of the logarithmic variable of market value will increase the return rate of stocks by 8.66\% before the stock crash. However, during and after the stock market crash, market capitalization's positive effect on stock returns dropped to about 3\%. The larger the net asset (owner's equity) is, the negative effect on stocks' return rate is shown in the regression, suggesting that a higher price/book (PB) ratio is likely to have a higher return, contrary to conventional wisdom. Each unit growth of the logarithmic variable of net assets in the current period will significantly reduce the return rate of stocks. Although the change of each unit of this variable only reduces the return rate of stocks by 1.5\% when the stock market crash occurs, it will have a negative impact of more than 5\% before and after the stock market crash. However, the price/book (PB) ratio is more critical in stocks' long–term performance and has a more significant impact on stock value decisions over the long term than over the short term.

In addition, the stock market's long–term investment theory suggests that companies with a low PE ratio may have more room to grow in the future. Thus, a low PE ratio requires as much net profit as possible at the same market value or a smaller market value at the same level of net profit. In fact, in this paper's short–term model, the effect of net profit on the rate of return is ambiguous. With the introduction of the quadratic term, the coefficient of net profit is significant only in the first and third periods. The main reason is that the financial market vibration caused by the stock market crash brings uncertain disturbance to the net profit of enterprises. Under the standardization premise, net profit's regression coefficient is also small compared with net assets and market value. The impact of standardized net profit varies in different periods. Since only 4 listed companies' standardized net assets were greater than the unit root (11.379) before the stock crash, the increase of net assets will generally reduce the expected return rate of stocks unless the company scale reaches a giant level. When a stock crash occurs, the effect of net assets is not significant. After the stock crash, only 3 listed companies' standardized net assets are greater than the unit root (18.8878). Therefore, the increase of net assets will generally increase stocks' return rate, unless the company scale reaches a very high level. The ranking of these giant companies has not changed dramatically in the short term.

However, each network cluster index has a more significant result for fitting the short–term return rate. The increase of degree will reduce the rate of return before the stock crash. Each additional node of each listed company is expected to reduce a single stock's return rate by 0.037\%. However, the increase of degree during the rescue period will increase the stocks' return rate, which is expected to decrease by 0.019\% during the stock crash but increase by 0.0015\% after the crash. After the stock market crash, the inhibitory effect of the density of the shareholding network on stocks' return rate is reduced to some extent. In addition to the degree, betweenness centrality and clustering coefficient may also affect stocks' return rate. Similar to the change of the degree's influence before and after the stock market crash, the positive effect of betweenness centrality degree on the return rate of stocks is weakened by the stock market crash. This variable's unit effect on stocks' return rate is 0.00000293, 0.00000132, and –0.000000321, respectively, in three periods, but its significance decreases continuously. The clustering coefficient's positive effect on the return rate of stocks is significant before and after the stock market crash, with coefficients of 0.4662 and 0.03245, which are basically in the same order of magnitude. However, during the stock market crash, the coefficient changes significantly and is not significant.

Among many indicators of centrality, betweenness centrality effectively measures the ability of each node as a hub. It measures the degree to which one node is located on the path between other nodes, which can be understood as particular social capital in a cross–shareholding network. A node with a high degree of betweenness centrality, such as a public company, may have considerable influence in a network due to its control of information transfer between other nodes. Therefore, once the listed company with a large betweenness centrality is removed from the network, it will cause a great interference to listed companies' equity network connectivity. These influential companies are located in the path of realizing the maximum number of two–point connectivity in the equity network. Therefore, these special connections may serve as vertical merger channels and horizontal competition in the future. Hub companies may act as middlemen, matching listed companies with lower customer acquisition costs. Therefore, enterprises with high betweenness centrality will act as the hub in the investment network before the stock market crash, and its existence has a significant effect on improving the return rate of stocks. However, due to the stock market crisis, the government and private measures to rescue the market increase the equity transactions, contributing to the increase in the number of connections in the whole network. Besides, the network structure gets more compact. The unique position of these hub listed companies will be suppressed and weakened, so the influence coefficient of the betweenness centrality becomes smaller, and its significance decreases.

The clustering coefficient is a particular Boolean variable that can be used to define the cluster's edge. In this paper, the average clustering coefficient of nodes is adopted, so this index is smoother than the Boolean variable. In the three groups, the large clustering coefficient will lead to an increase in stock return rates. In particular, the clustering before and after the stock market crash is more evident than that in the market turmoil, and the absolute value of the coefficient is larger and more significant. However, the visualization demonstrates that the clusters before the stock market crash are more diverse than the others, and the rescue decision and equity trading after the stock market crash connect more listed companies. The rescue of the market mainly comes from the influx of off–market funds. Equity link is mainly due to the changes in listed companies' portfolio by institutional investors and individual investors, i.e., the relationship between listed companies linked by the common ten shareholders. Therefore, before and after the stock market crash, the node relationship in each node's neighborhood is more stable. A larger clustering coefficient means that the listed companies surrounding a common listed company are more likely to have connections. The more the clustering coefficient is closer to 1means, the more likely the neighborhood of the listed company is to be complete, while the central company is more likely to exist in a network of clustering stably to resist the external shocks which is beneficial to improve the return on stocks. Therefore, when stable clustering is formed before and after a stock crash (rather than when it occurs), although the number of clusters displayed by modularization is different from the network structure, the clustering coefficient of clear clustering still has an apparent positive influence on the return rate of stocks.
By observing the three groups' modular classification, it can be found that the distribution of nodes in the cluster is not average. Four groups in the first group exceed 10\%, the sum ratio of 62.29\%. In the second group, the sum of the largest three groups accounts for 68.99\% of all nodes. The third group is more concentrated than the other two, with the largest group owning 52.52\% of the system's nodes.

\begin{table}[H]
  \centering
  \begin{threeparttable}
  \caption{The proportion of nodes in the modular group to the total nodes in system}
\label{ratio}
\begin{tabular}{llll}
\toprule
Modularity Class & 2015.03.01–2015.05.31 & 2015.06.01–2015.08.31 & 2015.09.01–201511.30 \\ \hline
0                & 16.15                 & 5.18                  &                      \\
1                & 3.21***               &                       & 10.26                \\
2                & 7.59                  & 18.91                 & 52.52                \\
3                & 5.91**                & 5.45                  & 36.99                \\
4                & 19.24***              & 12.3                  & 0.08                 \\
5                & 7.12                  & 4.99                  & 0.08***              \\
6                & 0.16                  & 37.78***              &                      \\
7                & 14.27                 & 5.57***               &                      \\
8                & 12.63***              & 9.82**                &                      \\
9                & 6.73***               &                       &                      \\
10               & 0.08***               &                       &                      \\
11               & 0.12                  &                       &                      \\
12               & 0.16                  &                       &                      \\
13               & 6.65*                 &                       &            \\    

\bottomrule
\end{tabular}
\begin{tablenotes}
\footnotesize
\item[]
Standard errors are in parentheses.\\
$ ^{***} p<.01, ^{**} p<.05, ^* p <.1$\\
\end{tablenotes}        
\end{threeparttable}       
\end{table}

Disturbances during the stock market crash make the change between 06.01. 2015 and 08.31, 2016 elusive. Therefore, the instrumental variable method is used to optimize the regression. Although the heteroscedasticity exists, the correlation coefficient matrix shows that the heteroscedasticity does not affect the coefficient estimation's unbiasedness. By comparing the least square method and the two–stage least square method, it can be proved that the efficiency improvement of net profit estimation requires the accuracy loss of the coefficient estimation of net assets. Considering that net assets are more suitable as a long–term indicator, and the accuracy loss of 2SLS estimation is not particularly large with the model fitting coefficient improving after re–estimation, the method of 2SLS is also a good model optimization.

However, the model used in this article still has much room for improvement. Future research can assign different weights to nodes according to the types of connections they establish. In small networks, edge weights will have a more prominent impact on the topology. For example, direct cross–shareholding connections between listed companies tend to have closer relationships and are more likely to conduct frequent business than listed companies linked by joint shareholders. Besides, although the top ten shareholders of the listed company all play a decisive role in the company's operation, share segmentation is still one of the indicators of the company's control power, so the shareholding ratio can also be used as the source of side weight assignment.

\subsection{Conclusions and Suggestions}
The paper introduces network topology into the cross–shareholding network. With the help of tools such as model optimization and regression analysis, it is confirmed that the characteristics of the shareholding network will significantly impact the return rate of stocks. Although the model estimation is not perfect in the period of risk propagation in the network, stocks' market performance and network position become influential factors affecting its return rate when the model disturbance is small.

The nature of network topology has a significant impact on the return rate of stocks, and its influence coefficient is still significant even in the particular period in the face of the stock crisis. The market value has a significant positive effect on the return rate of stocks, while the effect weaken sharply during and after the stock market crash. The negative effect of net assets on stocks' return rate are more prominent before and after the shock.

The increase of degree will reduce the rate of return before the stock crash. However, the increase of degree during the rescue period will increase the stocks' return rate, which is expected to decrease during the stock crash. Similar to the change of the degree's influence before and after the stock market crash, the positive effect of betweenness centrality degree on the return rate of stocks is weakened by the stock market crash.  Due to the stock market crisis, the government and private measures to rescue the market increase the equity transactions, contributing to the increase in the number of connections in the whole network.  Besides, the network structure gets more compact. When stable clustering is formed before and after a stock crash (rather than when it occurs), the clustering coefficient of clear clustering still has an apparent positive influence on the return rate of stocks.

Policymakers and institutional investors can predict which companies will have a decisive impact on risk contagion when potential risks come to optimize investment portfolios. For example, the collapse of the node enterprise with high betweenness centrality will cause more significant total loss to the system than the collapse of the peripheral enterprise. Enterprises with a higher degree may have a higher clustering coefficient and thus become the core of new clustering. Therefore, in the rescue process, the country can quickly find the most effective point to retard risk contagion. Although the cluster number of modularization is reduced after the stock market crash in China, in the longer term, the network of equity will once again become fragmented into different modules, rather than sticking together, as the connections created by the influx of external capital are likely to be temporary.

The analysis of cross–section data is only the beginning of verifying the validity of network analysis. If we want to better use the financial indicators of net assets and net profits, we need to establish long–term panel data and realize the evolution simulation of a dynamic networks. Although this paper tries to use the two–stage least square method to optimize the fitting of variables such as net profit in the regression equation, it still needs to dig deep into the financial market and shareholding network to obtain better instrumental variables. However, for the characteristics of network economics, the better fitting effect still requires empirical optimization of the model, addition of high–quality parameters with economic significance, and continuous iteration.

\nocite{*}
\bibliography{paper} 
\end{document}